\documentclass[aps,pra,twocolumn,showpacs]{revtex4}

\newcommand{\minus}[1]{ \mbox{$ (-1)^{#1} $}}

\newcommand{\ket}[1]{ \mbox{$|\, #1\,\rangle$}}

\newcommand{\ME}[3]{ \mbox{$\langle #1\,|\,#2\,|\,#3\rangle $} }

\newcommand{\MEred}[3]{ \mbox{$\langle #1\,||\,#2\,||\,#3\rangle $} }

\newcommand{\CGC}[6]{ \mbox{$ \left( #1 #2 , #3 #4\,|\, #5 #6 \right) $} }

\newcommand{\SechsJ}[6]{ \mbox{$ %
            \arraycolsep0.25ex %
            \left\{ \begin{array}{ccc} %
                       #1 & #2 & #3 \vspace{0.5ex}\\%
                       #4 & #5 & #6 %
                   \end{array} \right\} $} }

\def\sumint{\hbox{\small$\Sigma$}\kern-0.73em\int\kern.1em}

\usepackage{amsmath}

\usepackage{bm}
\usepackage{color}
\usepackage{graphics}
\usepackage{graphicx}

\begin{document}
\title{Quantum coherent control of the photo\-electron angular distribution in bichromatic ionization of atomic neon}

\author{E.~V.~Gryzlova$^1$, A.~N.~Grum-Grzhimailo$^1$, E.~I.~Staroselskaya$^{1,2}$, N.~Douguet$^{3,4}$, and K.~Bartschat$^3$}

\affiliation{$^1$Skobeltsyn Institute of Nuclear Physics, Lomonosov Moscow State University,
Moscow 119991, Russia}
\affiliation{$^2$Faculty of Physics, Lomonosov Moscow State University, Moscow 119991, Russia}
\affiliation{$^3$Department of Physics and Astronomy, Drake University, Des Moines, Iowa 50311, USA}
\affiliation{$^4$Department of Physics, University of Central Florida, Orlando, Florida 32816, USA}

\date{\today}

\begin{abstract}
We investigate the coherent control of the photo\-electron angular distribution in bichromatic atomic ionization. Neon is selected as
target since it is one of the most popular systems in current gas-phase experiments with free-electron lasers (FELSs).
In particular, we tackle practical questions, such as the role of the fine-structure splitting,
the pulse length, and the intensity. Time-dependent and stationary perturbation theory are employed, and we also solve the
time-dependent Schr\"{o}dinger equation in a single-active electron model. We consider neon ionized by a FEL pulse whose
fundamental frequency is in resonance with either $2p-3s$ or $2p-4s$ excitation. The contribution of the non\-resonant
two-photon process and its potential constructive or destructive role for quantum coherent control is investigated.
\end{abstract}

\maketitle

\section{Introduction}\label{sec:intro}

The manipulation of quantum interference is a powerful tool to demonstrate fundamental principles of quantum mechanics,
as well as to unravel the structure and dynamics of atoms, molecules, and clusters. The phase control of photo\-processes
has been the subject of numerous studies and reviews~\cite{Shapiro86,Shapiro00,Ehlotzky01,Brumer03,Shapiro2006,Astapenko06,Brif10}.
Quantum coherent-control approaches have been developed for such promising applications as control of chemical reactions or
biological changing~\cite{Zewail2000}, ultrafast and nonlinear optics~\cite{Nikolopoulos2005,Nabekawa2006},
and 4D ultrafast electron microscopy \cite{Zewail2010,Kwon2010},
to name just a few.

To observe quantum interference, one needs to create several (at least two) possible pathways leading
from a given initial state to the same final state.
This can be achieved in different ways, e.g., by using an intense field
(where $n$- and $(n+2)$-photon transitions interfere)~\cite{Schafer1992,Gao1998},
multi\-color fields (where different numbers of photons are needed to reach the same energy)~\cite{Swoboda2010,Petersson2016},
or short few-cycles pulses (whose spectral profile resembles a multicolor field)~\cite{Krausz09,Ranitovic2014,He2016}.
The combination of different effects is, of course, also possible.

Phase control for two-color setups, such as with $\omega$ and 2$\,\omega$, towards final states with equal parities
(e.g., $\omega+\omega+\omega+\omega$ and $2\,\omega+2\,\omega$) can already be seen in
the angle-integrated probability of the electron or ion yields~\cite{Nikolopoulos06,Sekikawa2008}. In contrast, mixing states of different
parities can only be observed in differential characteristics, such as the photo\-electron angular distribution (PAD).
Demonstrations of the interference between even- and odd-order processes have been reported previously,
including control over the PAD in
atomic~\cite{Muller1990,Schumacher1994,Yin1992,Baranova92,Schumacher1996} or molecular~\cite{Yin1995} photo\-ionization processes,
over the angular distribution of the products in molecular photodissociation~\cite{Sheehy1995},
as well as over the direction of electron emission and photocurrents in solids~\cite{Baranova91,Atanasov1996,Petek1997}.
The important difference between coherent photo\-ionization
of an atom and a molecule is that the latter does not possess spherical symmetry and, consequently, interference of waves
with opposite parities may be observed in both angle-integrated parameters (electron or cation yields) and the angular
distribution of the reaction products. For atoms, the use of static electric fields to mix states with different parity has
also been discussed~\cite{Manakov99,Bolovinos2008}.

Whereas different quantum coherent-control schemes have been developed and efficiently used in the optical and IR regions,
application of similar approaches at higher frequencies is hampered by the lack of coherence in typical
vacuum ultra\-violet (VUV) sources.
The recent implementation of a seeding scheme in the FEL FERMI provides high longitudinal coherence~\cite{Allaria2012}
and thus allowed to observe interference of two- and one-photon pathways in the PAD created by VUV radiation~\cite{Prince2016,Hartmann2016}.

A number of theoretical approaches have been developed to describe such experiments. Among these, the simplest ones
are based on non\-stationary perturbation theory.  They have the advantage of being able to present observables
in the form of amplitudes and phases associated with different channels, and under certain conditions allow
extraction of these amplitudes and phases~\cite{Nakajima2000,Wang01}. Nakajima~\cite{Nakajima2000}, for example,
applied this approach to the ionization of alkali elements, where the number of channels is very limited. For systems
where electron-electron correlations are important, for example noble-gas atoms, more sophisticated approaches are required.
Only recently the role of resonances in quantum coherent control was examined, especially regarding
the phase dependence of a pathway that passes through a resonance, demonstrating jumps~\cite{Anderson1992,Lee1998,Douguet2016},
Rabi oscillations leading to saturation of ionic or electronic yields~\cite{Fedorov2000,Nikolopoulos2015}, and evolution of
an autoionizing line~\cite{Buica2004,Lyras1999,Jimenez2014}.
Measurements of the phase lag were performed based on $\omega+2\,\omega$ ionization of excited barium
in the vicinity of an autoionizing state~\cite{Yamazaki07,Yamazaki07a}.

In a recent study~\cite{Douguet2017}, we considered coherent control of the PAD in
bi\-chromatic ionization passing through resonant excitation of the $(2p^53s)^1P$ inter\-mediate state in neon by the fundamental component of the radiation. In the present paper, we extend this work by considering also the actual case studied in the recent FERMI
experiment~\cite{Prince2016}, where one of the two $2p^54s$ states with total electronic angular momentum $J=1$ was used as the
inter\-mediate state while the other one, as well as the $J=1$ states of the $2p^53d$ configuration, also might have had an effect.
Specifically, we investigate the influence of various experimental parameters.

\begin{figure}
\includegraphics[width=\columnwidth]{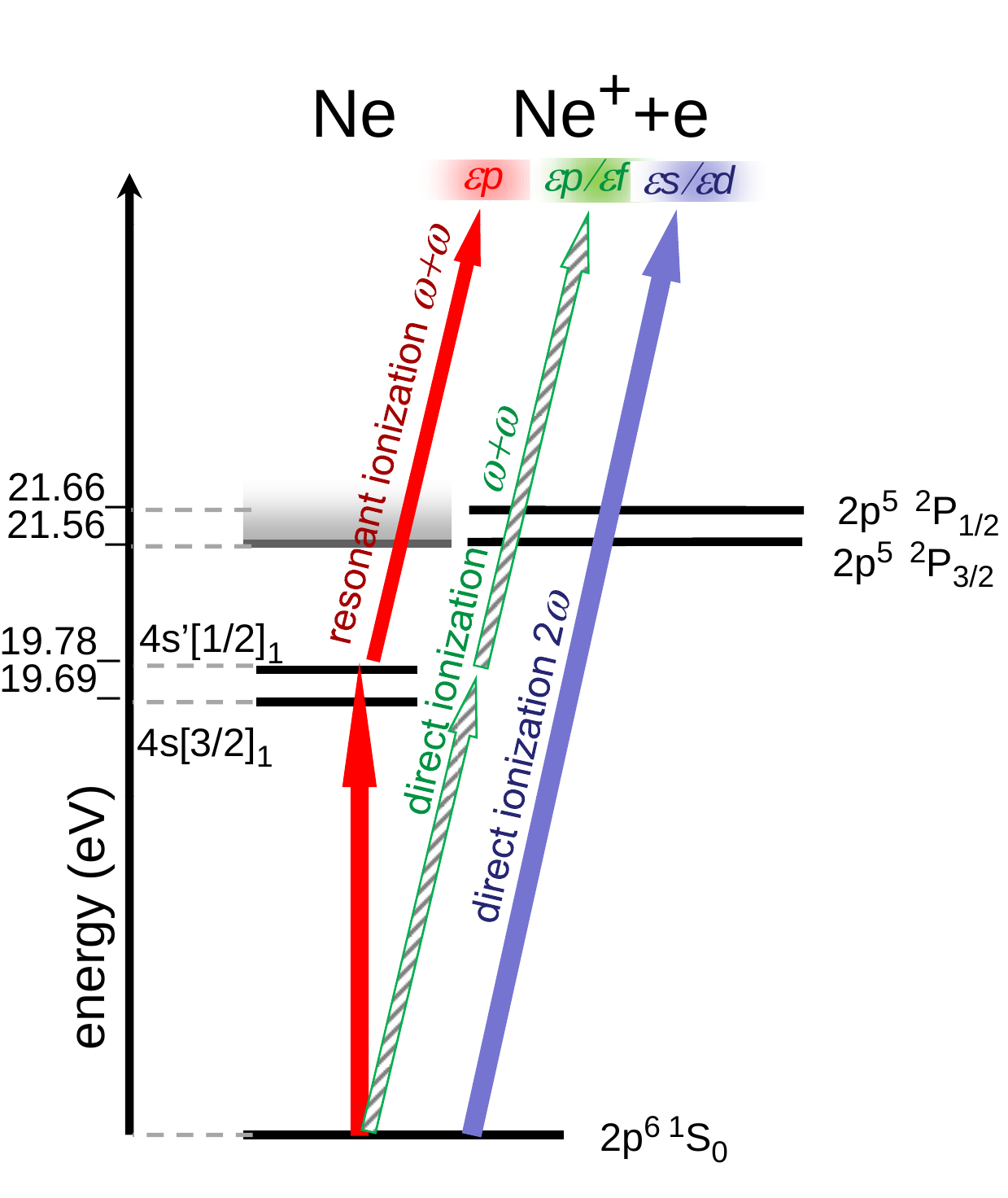}
\caption{
 Transition scheme for valence ionization of the neon ground state by bi\-chromatic radiation with fundamental harmonic energy~19.7~eV.}
\label{fig:fig1}
\end{figure}

For relatively small intensities and long pulses (more than about ten optical cycles),
at photo\-electron energy $2\,\omega-{\rm IP}$ (where IP denotes the ionization potential),
three ionization mechanisms by the bi\-chromatic field can be distinguished (cf.~Fig.~\ref{fig:fig1}):
(i)~single-photon ionization, (ii)~resonant two-photon ionization, and (iii) direct two-photon ionization.
The separation of the second and third mechanisms goes back to the very beginning of scattering theory and essentially
presents the difference between resonant and potential scattering amplitudes~\cite{Landau,Baz}.

Besides its practical interest, considering bi\-chromatic coherent control for resonant excitation of a $2p^54s$ state possesses a
few important features: Ne$\,(2p^54s)$ is much more correlated than the previously considered H$\,(2p)$~\cite{Grum2015}
and Ne$\,(2p^53s)$~\cite{Douguet2017} cases. Hence, it allows us to study the role of multi\-electron effects on coherent control.
Note also that the total spin of the target states with total electronic angular momentum $J=1$
is no longer a good quantum number, and thus the admixture of triplet states violates
the non\-relativistic selection rules. Furthermore, since the inter\-mediate state in this case is excited
less than the inter\-mediate states H$\,(2p)$ and Ne$\,(2p^53s)$, we can better disentangle resonant and direct transition amplitudes,
and hence their role in coherent control.

This manuscript is organized as follows. Section~\ref{sec:Theory} summarizes the basic theory and is followed by
the presentation of our results in Sect.~\ref{sec:Results}, where the role of various parameters is discussed
in individual subsections. Unless specified otherwise, atomic units are used throughout.

\section{Theory}\label{sec:Theory}

\subsection{General Considerations} \label{subsec:general}
The electric field of a bichromatic radiation is a combination of the fundamental ($\omega$) and second harmonic ($2\omega$)
frequencies, which we take in the form
\begin{equation} \label{eq:field}
\bm{E}(t)=\bm{e}
F(t)\left[\cos\omega t + \eta \cos (2\omega t+\varphi)\right]\, .
\end{equation}
Here $\bm{e}$ is the unit vector of the field polarization,
$F(t)=F_0\sin^2\left(\pi\, t/\textmd{T}\right)$ is the pulse envelope of duration $\textmd{T}$ containing
$N=\omega\,\textmd{T}/2\pi$ optical cycles, and $\eta$ is the amplitude ratio of the second and fundamental harmonics.
The intensity of these components may vary depending on the specific characteristics of the source.  Typically
the second harmonic intensity amounts to a few percent~\cite{Saldin1998} of the fundamental. For $N\gg 1$, it can be
shown by applying the rotating-wave approximation (RWA) that all observables depend only on the relative phase $\varphi$
but not on the individual carrier-envelope phases of the first and second harmonics.

In second-order perturbation theory~\cite{Messiah},
the differential ionization probability
can be expressed as
\begin{equation} \label{eq:rate}
\frac{d W}{d \Omega} = \frac{1}{2J_0+1} \sum_{M_f \mu M_0}
\left|  \eta U^{(1)}_{M_0,M_f \mu} + U^{(2)}_{M_0,M_f \mu} \right|^2 \,,
\end{equation}
where $U^{(1)}_{M_0,M_f \mu}$ is the first-order
amplitude for ionization by the second harmonic and
$U^{(2)}_{M_0,M_f \mu}$ is the second-order amplitude for two-photon ionization
by the fundamental; $M_0$, $M_f$, and $\mu$ are the magnetic quantum numbers of
the initial atomic state (with total electronic angular momentum $J_0$),
the residual ion (with~$J_f$),
and the photo\-electron, respectively.

In the electric dipole approximation,
\begin{equation} \label{eq:a1}
U^{(1)}_{M_0,M_f \mu} =-i \ME{\zeta_f J_f M_f, \bm{k} \mu^{(-)}}{\bm{e \hat{D}}}{J_0M_0} \, T^{(1)}
\end{equation}
and
\begin{eqnarray} \label{eq:a2}
U^{(2)}_{M_0,M_f \mu} = &-&\int  \hskip-5.5truemm \sum_{~~n}
\ME{\zeta_f J_f M_f, \bm{k} \mu^{(-)}}{\bm{e \hat{D}}}{\zeta_n J_n M_n}\nonumber\\
&\times&\ME{\zeta_n J_n M_n}{\bm{e \hat{D}}}{\alpha_0 J_0 M_0} \, T^{(2)}_{E_n} \,.
\end{eqnarray}
Here $\zeta_n$ specifies the quantum numbers
of the inter\-mediate states~$n$, while $\sumint$ denotes the infinite sum over all discrete states
and the integral over the continuum states.
The minus sign in the bra state indicates the proper
asymptotic form of the continuum wave function, $E_n$ is the
energy of the inter\-mediate virtual atomic state,
and the time-dependent factors $T^{(1)}$ and $T^{(2)}_{E_n}$ were described in
detail in Refs.~\cite{Grum2015,Douguet2016}.
The spherical component~$D_q$ of the electric dipole transition operator $\bm{\hat{D}}$
in the non\-relativistic long-wave\-length approximation is given by
\begin{equation} \label{eq:d}
D_q = \sqrt{\frac{4 \pi}{3}} \sum_p r_p Y_{1 q}(\theta_p, \phi_p),
\end{equation}
where the summation is taken over all atomic electrons,
$(r_p, \, \theta_p, \, \phi_p)$ are the electron spherical coordinates, and
$Y_{kq}(\theta, \phi)$ are spherical harmonics~\cite{Varshalovich}.

From now on we assume $J_0=0$. After some angular-momentum algebra~\cite{BGGK},
the amplitudes~(\ref{eq:a1}) and~(\ref{eq:a2})
for linearly polarized light ($q=0$, quantization axis $z \parallel \bm{e}$)
can be cast into the form
\begin{equation} \label{eq:fact}
U^{(\nu)}_{M_0,M_f \mu} = \delta_{M_0,0} \delta_{M_f, -\mu} U^{(\nu)}_{lj1} \,
\qquad \nu = 1,2 \,,
\end{equation}
where
\begin{equation} \label{eq:a12}
U^{(1)}_{lj1} = -i\frac{1}{\sqrt{3}}
i^{-l} e^{i \delta_l} \MEred{\zeta_f J_f, \varepsilon lj:J=1}{D}{0} \,T^{(1)} \,
\end{equation}
and
\begin{eqnarray} \label{eq:a22}
U^{(2)}_{ljJ} = &-&\frac{1}{\sqrt{3}\hat{J}}\CGC{1}{0}{1}{0}{J}{0} \,
i^{-l} e^{i \delta_l} \nonumber\\
&\times&\int  \hskip-5.5truemm \sum_{~~n}
\MEred{\zeta_f J_f, \varepsilon lj:J}{D}{\zeta_n J_n=1}\nonumber\\
&\times&\MEred{\zeta_n J_n=1}{D}{0} \, T_{E_n}^{(2)} \,.
\end{eqnarray}
Here $\varepsilon$ is the energy of the photoelectron, while $l$ and $j$ are its orbital and
total angular momentum, respectively.
We introduced $\hat{a} = \sqrt{2a+1}$ and the reduced matrix elements of the dipole
operator~\cite{Varshalovich},
and we used standard notations for the Clebsch-Gordan coefficients.
Below we will abbreviate  $-i\cdot i^{-l} e^{i \delta_l}\equiv e^{i \Delta_l^{(1)}}$ and
$-i^{-l} e^{i \delta_l}\equiv e^{i \Delta_l^{(2)}}$.

We can now write Eq.~(\ref{eq:rate}) as a sum of three terms:
\begin{equation} \label{eq:three}
\frac{d W}{d \Omega}  =   \frac{d W}{d \Omega} ^{( I)} +\frac{d W}{d \Omega}^{(II)} +\frac{d W}{d \Omega} ^{(III)}.
\end{equation}
In the above equation, the first and second terms are the ionization rates due to the second harmonic and the
first harmonic, respectively, while the third term is due to the interference between the two
paths.

Performing the summation over all projections of the angular momenta, transforming to the $LSJ$-coupling scheme,
and summing over the quantum
numbers $j, \, j', \, J_f$ (the last summation is performed incoherently), we obtain
\begin{eqnarray} \label{eq:s1}
\frac{d W}{d \Omega} ^{( I)} &=&|\eta|^2 \frac{\minus{L_f} }{4 \pi} \sum_{kll'} \hat{l} \hat{l}' \CGC{1}{0}{1}{0}{k}{0}
\CGC{l}{0}{l'}{0}{k}{0} \, \nonumber\\
&\times& \SechsJ{l}{l'}{k}{1}{1}{L_f}  e^{i(\Delta_l^{(1)} - \Delta_{l'}^{(1)})}
D^{(0)}_{0 \rightarrow L_f l, 1} D^{(0) \, \ast}_{0 \rightarrow L_f l', 1} \, \nonumber\\
&\times& P_k(\cos \vartheta) \,.
\end{eqnarray}
where $D^{(0)}_{0 \rightarrow L_f l, 1} = \MEred{\zeta_f L_f, l: L=1}{D}{L_0 = 0}$ and
$P_k(x)$ denotes a Legendre polynomial.
The superscript `$(0)$' indicates the conserved total spin $S_0 = S = 0$.  In principle,
the electron wave functions can depend on~$S$.
We also introduced the conventional notation for $6j$-symbols.
Note that the angle $\vartheta$ is counted from the direction of the electric field vector.

We now assume that the inter\-mediate states in the second-order amplitude are described in the
inter\-mediate-coupling scheme. For the $n$-th state we have
\begin{equation} \label{eq:cp}
\ket{\zeta_n J_n} = \sum_{L_n S_n} \alpha_{L_n S_n}^{\zeta_n} \ket{\zeta_n L_n S_n J_n} \,.
\end{equation}
Performing similar summations as for $(I)$, we obtain
\begin{eqnarray} \label{eq:s2}
\frac{d W}{d \Omega} ^{( III)} \hspace{-0.5truecm} &=&  \eta\, \frac{\minus{L_f}}{2\pi \sqrt{3}}
\sum_{ll'L'k} \hat{l} \hat{l}'
\CGC{l}{0}{l'}{0}{k}{0}\nonumber\\
&\times&\CGC{1}{0}{L'}{0}{k}{0}\CGC{1}{0}{1}{0}{L'}{0}
\SechsJ{l}{l'}{k}{L'}{1}{L_f} \nonumber\\
&\times&{\rm Re}\Bigg[e^{i(\Delta_l^{(1)} - \delta_{l'}^{(2)})} D^{(0)}_{0 \rightarrow L_fl,1}T^{(1)} \nonumber \\
&\times& \left( \sum_n \left| \alpha_{10}^{\zeta_n} \right|^2 T^{(2)}_{E_n} \,
 D^{(0) \, \ast}_{n,1 \rightarrow L_f l',L'} D_{0 \rightarrow n,1}^{(0) \, \ast} \right)\Bigg] \, \nonumber\\
&\times& P_k(\cos \vartheta) \,,
\end{eqnarray}
where \hbox{$D^{(0)}_{n,1 \rightarrow L_f l',L'} = \MEred{\zeta_f L_f,l':L'}{D^{(S=0)}}{\zeta_n L_n=1}$},
\hbox{$D^{(0)}_{0 \rightarrow n,1} = \MEred{\zeta_n L_n=1}{D^{(S_n = 0)}}{L_0=0}$}, and
$\alpha_{10}^{\zeta_n} \equiv \alpha_{L_n=1, S_n= 0}^{\zeta_n}$.
In the RWA, $T^{(1)}T^{(2)*}_{E_n}\sim\exp(-i\varphi)$.
Finally, for the second term in Eq.~(\ref{eq:three}), we obtain
\begin{widetext}
\begin{eqnarray} \label{eq:s3}
\frac{d W}{d \Omega} ^{( II)} &  = & \frac{\minus{L_f}}{4\pi}\sum_{ll'JJ'k} \hat{l} \hat{l}' \hat{J} \hat{J}'
\CGC{l}{0}{l'}{0}{k}{0} \CGC{1}{0}{1}{0}{J}{0} \CGC{1}{0}{1}{0}{J'}{0}
\CGC{J}{0}{J'}{0}{k}{0}e^{i(\Delta_l^{(2)} - \Delta_{l'}^{(2)})}\nonumber\\
&  & \times\sum_{SLL'} \minus{S+L+L'} \hat{L} \hat{L}' \SechsJ{l}{l'}{k}{L'}{L}{L_f} \SechsJ{L'}{L}{k}{J}{J'}{S}
\left(
\sum_{n, L_n} \alpha_{L_nS}^{\zeta_n} \alpha_{10}^{\zeta_n \, \ast} T^{(2)}_{E_n}
\SechsJ{S}{L}{J}{1}{1}{L_n} D^{(S)}_{n, L_n \rightarrow L_f l, L} D^{(0)}_{0 \rightarrow n,1} \right)\nonumber\\
&  & \times \left(
\sum_{n', L_{n'}} \alpha_{L_{n'} S}^{\zeta_{n'} \, \ast}
\alpha_{10}^{\zeta_{n'}} T^{(2)}_{E_{n'}}
\SechsJ{S}{L'}{J'}{1}{1}{L_{n'}} D^{(S) \, \ast}_{n', L_{n'} \rightarrow L_f l', L'}
 D^{(0) \, \ast}_{0 \rightarrow n',1} \right)
P_k(\cos \vartheta) \,.
\end{eqnarray}
\end{widetext}

Equations~(\ref{eq:s1})-(\ref{eq:s3}) allow us to present the PAD~(\ref{eq:three})
in the conventional form
\begin{equation} \label{eq:PAD}
W(\vartheta) \equiv \frac{d W}{d \Omega} =
\frac{W_0}{4 \pi} \left( 1 + \sum_{k=1}^4\beta_k P_k(\cos \vartheta) \right),
\end{equation}
in terms of the anisotropy parameters~$\beta_k$ and the angle-integrated ionization probability~$W_0$.
In second-order perturbation theory, the maximum rank is $k=4$.
From parity conservation and the properties of the Clebsch-Gordan coefficients,
it follows that (\ref{eq:s1}) and (\ref{eq:s3}) give contributions to the total
ionization probability and the even-rank asymmetry parameters,
while (\ref{eq:s2}) contributes to the odd-rank asymmetry parameters.
It is worth noting that, in contrast to Eq.~(\ref{eq:s3}), no term with $S\neq0$
appears in Eq.~(\ref{eq:s2}). Consequently, there is no interference
between singlet and triplet final states of the $e + {\rm Ne}^+$ scattering system.

The presence of odd-rank terms ($k=1, \, 3$) in~(\ref{eq:PAD}) means that the
forward-backward symmetry of the PAD along the
direction of the polarization vector is broken.  Specifically, the normalized
difference in the signals is given by
\begin{equation} \label{eq:Asym}
A(0) = \frac{W(0)-W(\pi)}{W(0)+W(\pi)}\, .
\end{equation}
The forward-backward asymmetry~(\ref{eq:Asym}) is a function of many parameters associated with the irradiating
beam, including the fundamental angular frequency~$\omega$, the peak intensities
(absolute and relative) of the
fundamental, $I=F_0^2/4$, and the admixture of the second harmonic $\eta^2 I$,
their relative phase $\varphi$, and the number of optical cycles~$N$, i.e., the length of the pulse.

The asymmetry~(\ref{eq:Asym}), as a function of the relative phase of the harmonics~$\varphi$,
can be presented in the form
\begin{equation} \label{eq:Asym1}
A(0)=A^{(m)}\cos(\varphi-\varphi_m)
\end{equation}
with the amplitude of the oscillations $A^{(m)}$ and the phase~$\varphi_m$,
where the asymmetry reaches its maximum value. In an actual experiment, such as that reported in Ref.~\cite{Prince2016},
changing the frequency near an inter\-mediate resonance may produce
a jump in the phase~$\varphi_m$.

The anisotropy parameters~$\beta_k$ with odd $k$ are described by equations similar to~(\ref{eq:Asym1})
with their own amplitudes and phases instead of $A^{(m)}$ and $\varphi_m$, respectively.
This increases the number of independent measurable parameters and raises the question of the possibility
of ``complete'' experiment~\cite{Kleinpoppen13}.

\subsection{Computational models}\label{subsec:computational}

We employ three different computational models to describe bichromatic ionization of neon, namely:
(i)~Time-dependent perturbation theory (PT);
(ii)~Perturbation theory for an infinitely long ``pulse'' (PT$_{\infty}$);
(iii) Solution of the time-dependent Schr\"{o}dinger equation (TDSE).
Our integration schemes for the TDSE have been described in detail
elsewhere~\cite{Abeln2010,GGKB2013,Ivanov2014}. The variational principle realized in our  PT$_{\infty}$ was
proposed in~\cite{Orel88,Gao1989}, and this approach was recently tested for the PADs
in~\cite{Staroselskaya2015}.

\begin{table*}[htb]
\caption{\label{tab:Table1} Excitation energies (in eV), mixing coefficients, and radial electric-dipole
matrix elements (in a.u.) for selected states in neon, identified in the $jlK$-coupling scheme.}
\begin{tabular}{l c c c c c c}
\hline
   & \multicolumn{3}{c}{Excitation energy}& Leading terms in MCHF & \multicolumn{2}{c}{Matrix elements}  \\
\cline{2-4} \cline{6-7}
State & exp & TDSE & HS & configuration mixing & MCHF &  TDSE	 \\\hline\hline
 $2p^5(^2\!P_{3/2})3s[3/2]_1$ & 16.67 &- &-& $-0.15|2p^53s \,^1\!P\rangle+0.98|2p^53s \,^3\!P\rangle$ & 0.09 & - \\
 $2p^5(^2\!P_{1/2})3s'[1/2]_1$ & 16.84 &16.35 &15.48&$0.98|2p^53s \, ^1\!P\rangle+0.15|2p^53s \, ^3\!P\rangle$ & -0.59 & -0.63 \\
 $2p^5(^2\!P_{3/2})4s[3/2]_1$ & 19.68 &- &-& $-0.58|2p^54s \,^1\!P\rangle+0.81|2p^54s\,^3\!P\rangle$ & 0.13 & -  \\
 $2p^5(^2\!P_{1/2})4s'[1/2]_1$ & 19.78 &19.29 &18.25& $0.81|2p^54s\,^1\!P\rangle+0.58|2p^54s\,^3\!P\rangle$ & -0.18 & -0.24 \\
 $2p^5(^2\!P_{3/2})3d[1/2]_1$ & 20.03 &- &-& $0.38|2p^53d\,^1\!P\rangle-0.91|2p^53d\,^3\!P\rangle+0.12|2p^53d\,^3\!D\rangle$ &
  0.08 & -  \\
 $2p^5(^2\!P_{3/2})3d[3/2]_1$ & 20.04 &19.62 &18.58& $0.77|2p^53d\,^1\!P\rangle+0.24|2p^53d\,^3\!P\rangle-0.58|2p^53d\,^3\!D\rangle$ &
 0.16 & 0.27 \\
 $2p^5(^2\!P_{1/2})3d[3/2]_1$ & 20.14 &- &-& $0.50|2p^53d \,^1\!P\rangle+0.32|2p^53d \,^3\!P\rangle+0.80|2p^53d \,^3\!D\rangle$ &
 0.10 & - \\\hline
\end{tabular}
\end{table*}

In the first approach (PT) we employ wave\-functions calculated with the Multi-Configuration Hartree-Fock (MCHF) code
described in~\cite{Froese97}. We froze the $1s,2s,2p$ orbitals obtained in a self-consistent calculation of
the Ne~$(1s^22s^22p^{6})^1S$ ground state,
then found the $3s,4s,3d$ orbitals optimized on the term-averaged energies of the $1s^22s^22p^5nl$ configurations,
and finally mixed all these configurations and $LS$ terms in the semi\-relativistic Breit-Pauli approach
to obtain the configuration and term-mixing coefficients.
For the seven Ne $2p^54s$ and $2p^53d$
states with $J_n=1$, the latter turned out to be close
to the coefficients obtained in the pure $jlK$ coupling scheme.
The calculated excitation energies of these states
in the MCHF model agree to better than 0.5\% with the experimental
ones given in  Table~\ref{tab:Table1}. Only the above seven excited states
were included explicitly.  All other inter\-mediate states, including the continuum,
were incorporated within the single-active electron (SAE) approximation in a model potential (see below).

The PT$_{\infty}$ and TDSE models are both non\-relativistic and employ SAE potentials
to describe the neon atom. Specifically, the PT$_{\infty}$ model is based on the Herman-Skillman
potential, while the TDSE model obtains its one-electron orbitals from the bound states calculated
in the Hartree potential formed with the Hartree-Fock orbitals of the $1s^2 2s^2 2p^5$ ionic core.
While these structure models are inferior to the MCHF calculation outlined above, both PT$_{\infty}$
and TDSE account for all excited and continuum states, the former within the second-order perturbation theory.

The ionization continuum of neon neither contains Cooper minima nor resonances in the energy region of interest.
All three approaches yield similar scattering phases and photo\-ionization amplitudes, as shown in Ref.~\cite{Douguet2017}.
On the other hand, the bound-bound matrix elements differ significantly.
This is illustrated in the last two columns of Table~\ref{tab:Table1}
for the MCHF and TDSE models.  Note also that the latter only account for the state with dominant $^1P$ character.
These differences are not due to significantly
different one-electron orbitals, but rather due to
the configuration mixing caused by the spin-orbit interaction that
splits the Ne$^+(2p^5)^2P_{J_f}$ ionic core. The experimental
ionization potentials are $21.56$ ($J_f=3/2$) and $21.66$ ($J_f=1/2$), while the corresponding
potentials in the TDSE and HS models (not split) are $21.16$~eV and $20.00$~eV, respectively.

In order to reduce the complexity of notation, we will now refer to the various states as follows:
$2p^5(^2P_{3/2})3s[3/2]_1 \equiv 3s$; $2p^5(^2P_{1/2})3s'[1/2]_1 \equiv 3s'$;
\hbox{$2p^5(^2P_{3/2})4s[3/2]_1 \equiv 4s$}; \hbox{$2p^5(^2P_{1/2})4s'[1/2]_1 \equiv 4s'$}.
Furthermore, we will group the three states $2p^5(^2P_{3/2,1/2})3d[3/2,1/2]_1$ together and refer to them as~$3d$.

\section{Results and discussion}\label{sec:Results}

This section presents our main results.  Particular emphasis is placed on elements that
are shown to have significant effects on the observed outcome:  the fundamental frequency, the pulse duration,
and the intensity, as well as the intensity ratio between the harmonics. In a theoretical treatment,
it is convenient to investigate each characteristic as function of
the fundamental frequency~$\omega$ while keeping all other pulse parameters constant.

In the TDSE and PT calculations presented below, we used pulses containing an integer number $N$ of
optical cycles, and hence the pulse duration is slightly different for different photon energies.
Nevertheless, for the long pulses considered in this study, this effect is negligible.
\subsection{Effects of the angular resolution}\label{subsec:averaging}

\begin{figure}[b]
\includegraphics[width=\columnwidth]{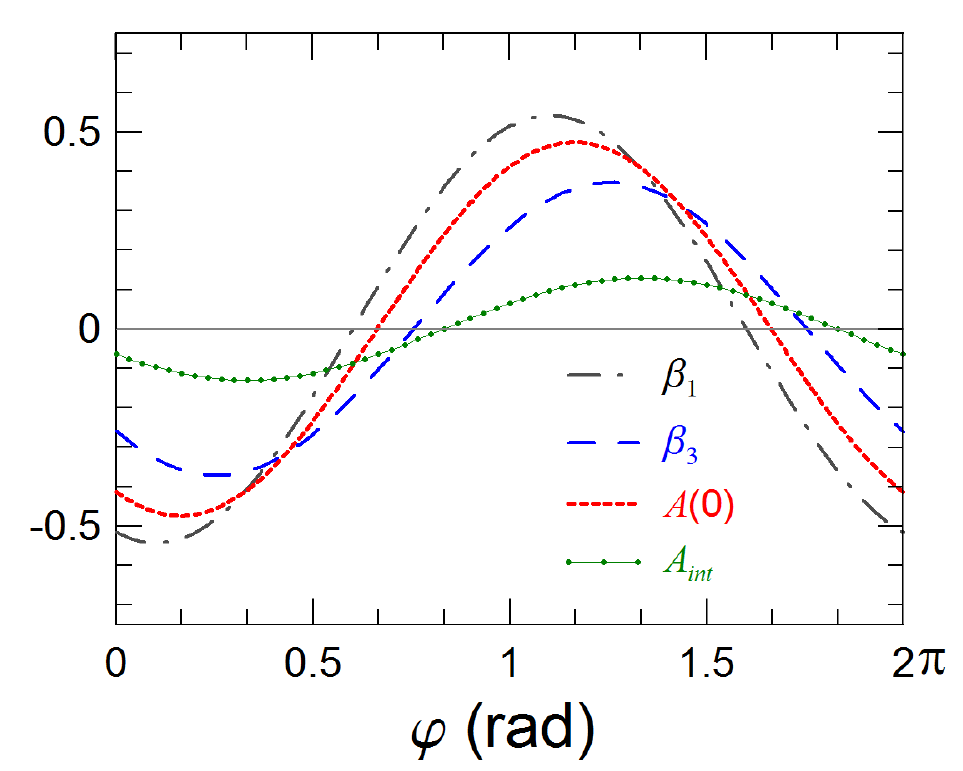}
\caption{
Anisotropy parameters and forward-backward asymmetry as function of the phase between the fundamental
and the second harmonic for a central photon energy of 19.78 eV, as obtained in the PT model.
The asymmetry~$A(0)$, Eq.~(\ref{eq:Asym}), for a perfect angular resolution is shown together
with the integral asymmetry defined in Eq.~(\ref{eq:Aint}).}
\label{fig:fig2}
\end{figure}

Figure~\ref{fig:fig2} presents PT results for the anisotropy parameters~$\beta_1$ and $\beta_3$,
as well as $A(0)$ as function of the phase between the harmonics at fixed mean photon energy.
In the actual experimental setup, however, the asymmetry $A(0)$ is determined after convolution
over some finite solid angle of the detected electrons.
Figure~\ref{fig:fig2} shows the extreme cases of ideal angular resolution and the angular
distribution integrated over each hemisphere (integral asymmetry). The latter was
determined experimentally in~\cite{Prince2016}.

\begin{figure*}[t]
\includegraphics[width=\textwidth]{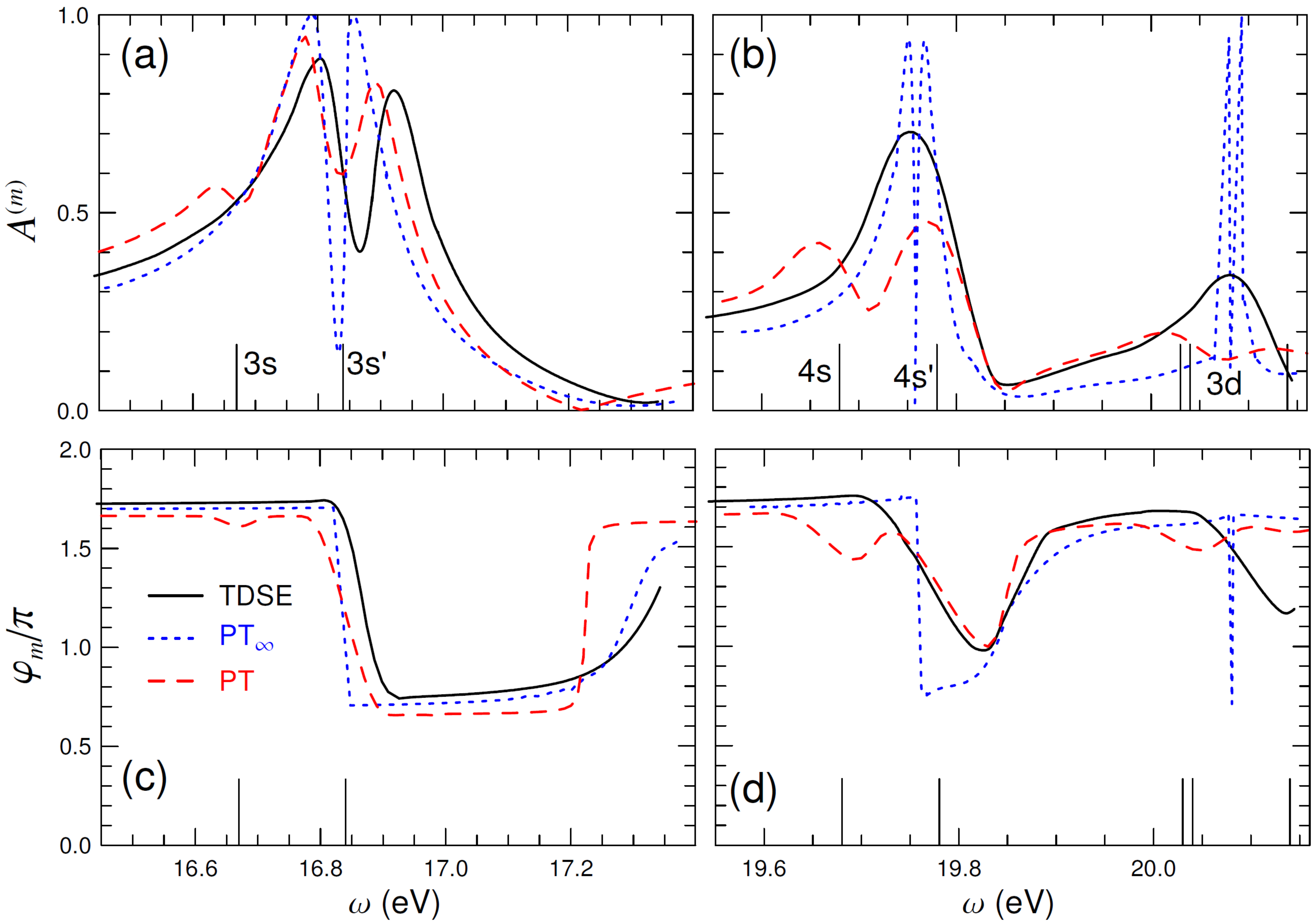}
  \caption{
 Maximal forward-backward asymmetry of the PAD and the corresponding phase as
 function of the fundamental photon frequency in the region of the $3s,3s'$ resonances (panels (a) and~(c)),
 and in the region of the $4s,4s'$, and $3d$ resonances (panels (b) and~(d)).
 The black solid lines correspond to the TDSE calculation, the red dash-dotted to PT,
 and the blue dashed to PT$_{\infty}$. The pulse parameters are $I=10^{12}\textrm{W/cm}^2$, $N=500$, $\eta=0.1$.
 The energies used in all theoretical curves were rescaled to match the term-averaged experimental resonance energy.}
  \label{fig:fig3}
\end{figure*}

The integral asymmetry is defined as
\begin{equation} \label{eq:Aint}
A_{int} = \frac{ \int_0^{\pi/2} W(\vartheta) \sin \vartheta d \vartheta -
\int_{\pi/2}^{\pi} W(\vartheta) \sin \vartheta d \vartheta}{\int_0^{\pi} W(\vartheta) \sin \vartheta d \vartheta}\,.
\end{equation}
Note that the result of the angular integration is not simply a reduction of the amplitude.
In fact, even the phase between the harmonics $\varphi_m$ (see Eq.~(\ref{eq:Asym1})),
corresponding to the maximum asymmetry, changes for different experimental conditions.
The amplitude of the forward-backward asymmetry $A(0)$ reaches $A^{(m)} \approx 0.50$ and
is achieved at $\varphi_m \approx 7\pi/6$, while the amplitude of the integral
asymmetry $A_{int}$ reaches only $A^{(m)} \approx 0.12$, and its maximum is
shifted to  $\varphi_m \approx 4\pi/3$.

\subsection{Dependence of the parameters on the fundamental frequency and pulse duration}\label{subsec:frequency}

We now consider variations in the PAD as a function of the fundamental frequency.
Figure~\ref{fig:fig3} shows the maximum asymmetry $A^{(m)}$ and the corresponding phase $\varphi_m$
as function of the photon energy.
We already predicted~\cite{Grum2015,Douguet2016,Douguet2017} that,
within PT$_{\infty}$, the asymmetry always reaches its maximum value on
both sides of the resonance, while the width of the structure depends on~$\eta$ and the ratio of
the two-photon and one-photon ionization amplitudes.
The former feature is seen in panels (a) and (b) of Fig.~\ref{fig:fig3}.

It is convenient to start the discussion in the vicinity
of the $3s'$ resonance at 16.84 eV~with dominant $^1\!P$ character (Figs.~\ref{fig:fig3}a,c).
We found that the real part of the two-photon amplitude near the resonance is
very large, and it abruptly changes sign when the photon energy crosses the resonance.
As a result, even though the two-photon and one-photon contributions
are comparable close to the resonance and hence cause strong interference and a significant
asymmetry, $A^{(m)}$ drops quickly from large values with
a corresponding jump of the phase~$\varphi_m$ by~$\pi$.
Once the jump is completed, $A^{(m)}$ goes back up.

This effect is seen in all three calculations, as displayed in panels~(a) and (c) of Fig.~\ref{fig:fig3}.  Due to the
finite width of the pulse, the changes are less rapid in PT and TDSE than in PT$_{\infty}$, but the
spectral width of a pulse with $N=500$ optical cycles ($\Delta \omega \approx 60$~meV)
is sufficiently small for the PT and TDSE calculations to essentially resolve these structures.

In contrast to PT$_{\infty}$ and TDSE, the PT model accounts for
spin-mixing in the inter\-mediate states.
The small feature near 16.67~eV is caused by the singlet admixture to the predominantly triplet $3s$ state, but
this detail is not expected to be resolvable in a currently feasible experiment.
Overall, therefore, the agreement between the predictions from the three models is good.

Next, we note that the two-photon amplitudes change sign also between the resonances~\cite{Bebb1966}.
We begin with the region between $3s'$ and $4s'$, i.e., the states that are ``seen'' in all
three models. In this energy region from about 17~eV to 19.5~eV (not shown), the phase $\varphi_m$
changes again by almost~$\pi$, but much more gradually than near the resonance.
Furthermore, $A^{(m)}$ is small (less than about~0.2), because
the interference term in Eq.~(\ref{eq:three}) is much smaller than the first one-photon term.

Increasing the photon energy further and looking at panels~(b) and~(d) of Fig.~\ref{fig:fig3},
we notice that the phase in PT$_{\infty}$ changes by~$\pi$ just below the $4s'$ and the highest of the $3d$ resonances.
In $LS$-coupling, the latter is the state with $^1P$ character in this manifold, and hence it is
the only one entering the PT$_{\infty}$ and TDSE models. For both the $4s'$ and the highest $3d$ state,
the phase jumps occur over such a small energy interval that only PT$_\infty$ can resolve them
and the associated double-peak structures in the maximum asymmetry.  The TDSE model, on the other hand,
convolves the structure with the bandwidth of the pulse.  Apart from that,
however, there is again satisfactory agreement between PT$_\infty$ and TDSE.

In contrast to the energy region around the $3s'$ resonance,
the PT results differ substantially
from PT$_\infty$, and hence also from TDSE, near the $4s$, $4s'$, and $3d$ resonances
(cf.\ Figs.~\ref{fig:fig3}b,d).  This is again due to the fact that PT as a
semi\-relativistic theory can ``see'' both the $4s$ and $4s'$ states, as well as all three states of
the $3d$ manifold. Since the $4s$ and $4s'$ states both contain very substantial singlet and triplet
components (cf.\ Table~\ref{tab:Table1}), it is not surprising that
structures due to the $4s$ state are clearly visible in the PT predictions now, in contrast to
the $3s$ state discussed above.  Regarding the $3d$ states, the pulse width is too large to resolve
them properly.

Although we concentrate on resonant excitation of the $ns$-states, it is important to note that the \hbox{$f$-wave}
actually dominates the \hbox{$p$-wave} over a broad energy region, as soon as the detuning $|\Delta|=|E_n-\omega| >0.1$~eV.
For our specific case in neon, the $f$-channel is due to the potential part of the transition amplitude,
rather than the resonant part~\cite{Landau,Baz}.
The importance of the $f$-wave has, indeed, posed problems in previous calculations~\cite{Moccia1983}.
The restricted PT employed in our
previous work~\cite{Douguet2017} also clearly underestimated the role of two-photon transitions into the \hbox{$f$-wave}.
This channel basically determines the zeros of the asymmetry at $\omega \approx 17.7$~eV and $\omega \approx 19.85$~eV, where it
plays the role of destructive interference. On the other hand, the \hbox{$f$-wave} ionization channel acts constructively
below the $3s'$ and $4s'$ states, thereby leading to a noticeable asymmetry.

\subsection{Optimal conditions for the realization of quantum control}\label{subsec:QC}

\begin{figure}[b]
\includegraphics[width=\columnwidth]{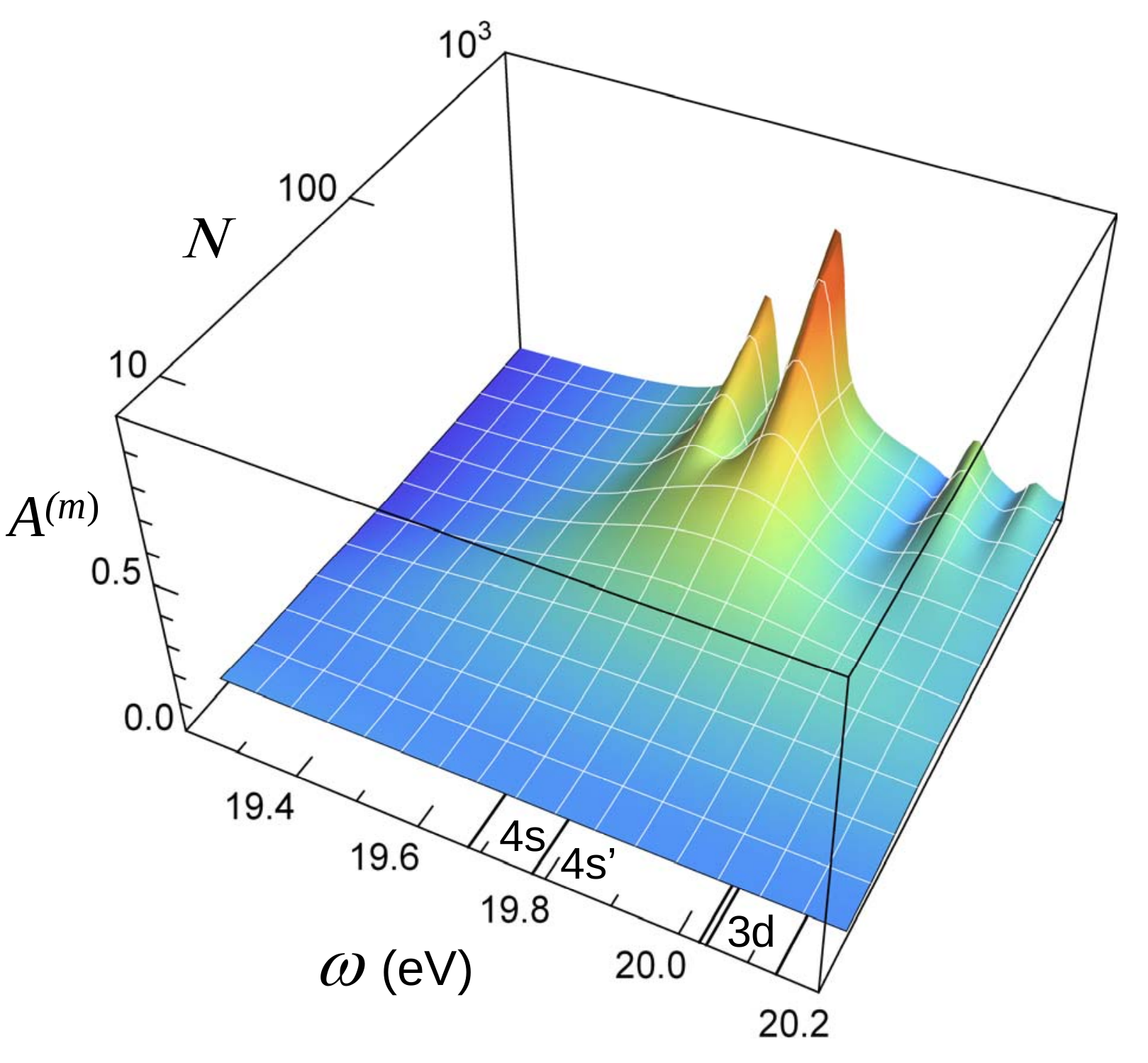}
 \caption{
 Maximal asymmetry as function of photon energy and pulse duration for $I=10^{12}$ W/cm$^2$
 and $\eta = 0.1$.}
\label{fig:fig4}
\end{figure}

In the scheme proposed in~\cite{Prince2016}, quantum coherent control of the PAD is achieved by
tuning the phase  $\varphi$ (via a time delay) between the fundamental and the second harmonic.
Whether or not the scheme is experimentally realizable
depends critically on the maximal PAD asymmetry at a given set of pulse parameters,
such as $\omega$, $\eta$, $N$, and the peak intensity $I$. If the asymmetry
is very small, changing the phase will only modify it a little. If the maximal
asymmetry is close to unity, however, one can completely change the direction of photo\-electron emission.

The asymmetry of the PAD builds up during the pulse as the result of many cycles, i.e., repetitions
of the field strength (just slowly modified by the envelope function) and its direction.
Figure~\ref{fig:fig4} shows the maximum asymmetry as a function of the photon energy and pulse duration.
Not surprisingly, the longer the pulse is, the narrower and spikier is the structure of the
asymmetry in the energy domain. For the case at hand, the pulse should contain at least 100 optical
cycles to efficiently form the interference pattern needed for a clearly detectable asymmetry.

Beyond the pulse duration, another important parameter to effectively observe coherent control of the PAD
is the amplitude ratio~$\eta$ defined in Eq.~(\ref{eq:field}), i.e., the admixture of the second harmonic.
If PT is valid and an inter\-mediate state is isolated,
there are two possible regimes. In the first one, increasing~$\eta$ narrows
the profile of the odd-rank anisotropy parameters~$\beta_1$ and~$\beta_3$, and hence of the asymmetry
(see Eqs.~(23) and (30) of~\cite{Grum2015}). In the second one, increasing~$\eta$ reduces the
asymmetry amplitude. The transition between the two regimes occurs when the width of the asymmetry profile
becomes smaller than the spectral width of the pulse.

\begin{figure}
\includegraphics[width=0.95\columnwidth]{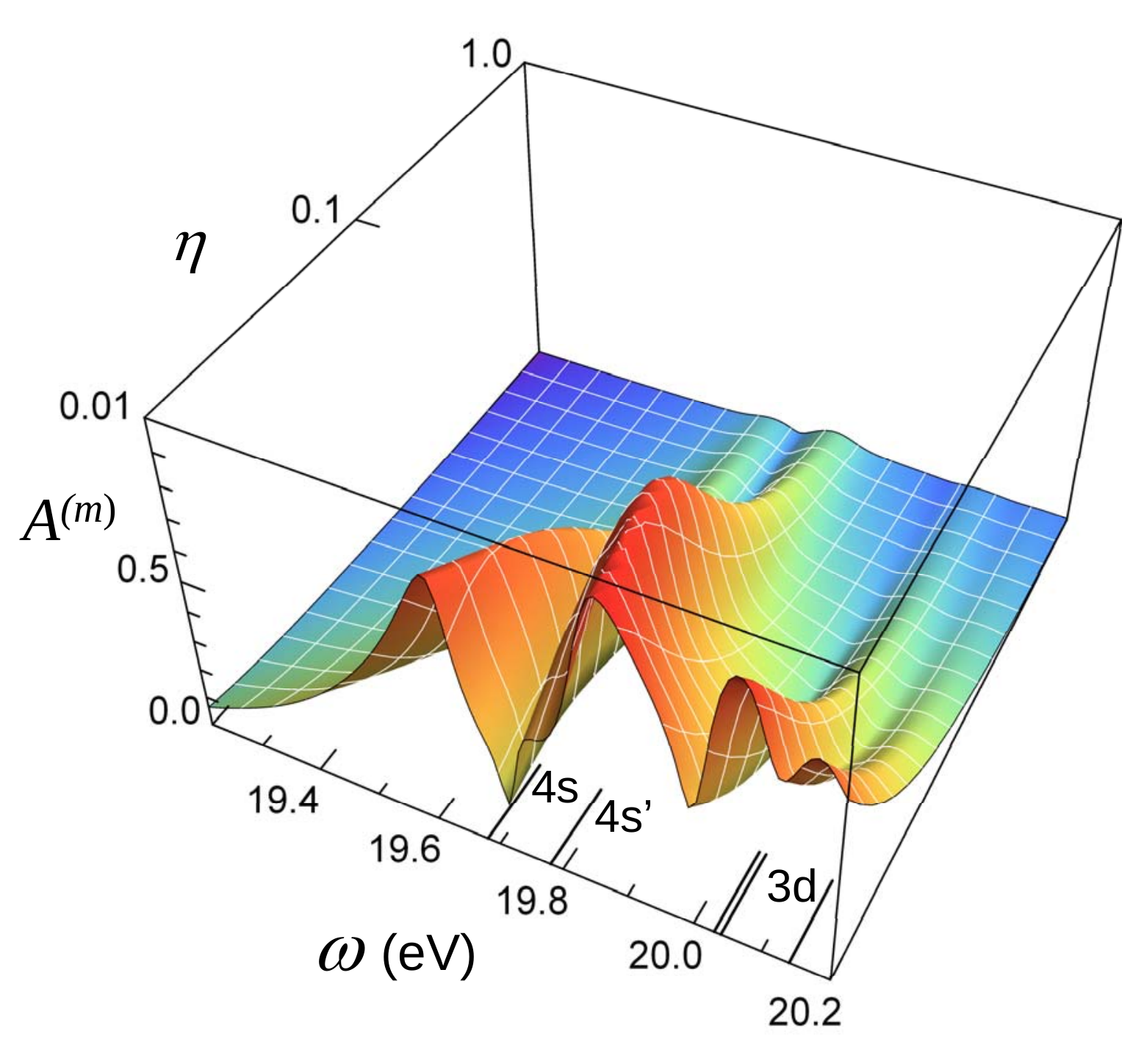}
\caption{Maximal asymmetry as function of the fundamental
energy $\omega$ and the ratio~$\eta$ of the second-harmonic strength
for $I=10^{12}$ W/cm$^2$ and $N = 500$.}
\label{fig:fig5}
\end{figure}

Choosing, for example, our pulse parameters as
$I=10^{12}$ W/cm$^2$ and $N=500$, we observe the first regime for $3s'$ and the transition
from the first to the second regime for $4s'$ (cf.\ \hbox{Fig.~\ref{fig:fig5}}).
Thus, whereas we can realize coherent control for various~$\eta$ values for $3s'$,
(from 0.001 to 1, see fig.~(9b) in \cite{Douguet2017}) our models predict optimal values of $\eta=0.02$ (PT) and $\eta=0.04$ (TDSE) for $4s'$.
Also note that there are no good conditions for observing significant interference near the lower $4s$ component:
If $\eta$ is large enough to resolve $4s$ from $4s'$, the interference is
already too weak, and it becomes weaker with increasing~$\eta$. Increasing the peak intensity~$I$
is effectively equivalent to decreasing~$\eta^2$, i.e., decreasing the relative intensity
of the second harmonic. Keeping the ratio $\sqrt{I}/\eta$ constant, therefore, one
can expect approximately the same asymmetry for different intensities.

The situation changes drastically when PT is no longer valid. Then
Rabi oscillations between the ground and the inter\-mediate states arise, and
the Autler-Townes splitting caused by the strong fundamental decreases the overlap
of the signals produced by the two harmonics.  This, in turn, reduces the asymmetry,
and thus prevents effective quantum control of the system.

\begin{figure}
 \includegraphics[width=\columnwidth]{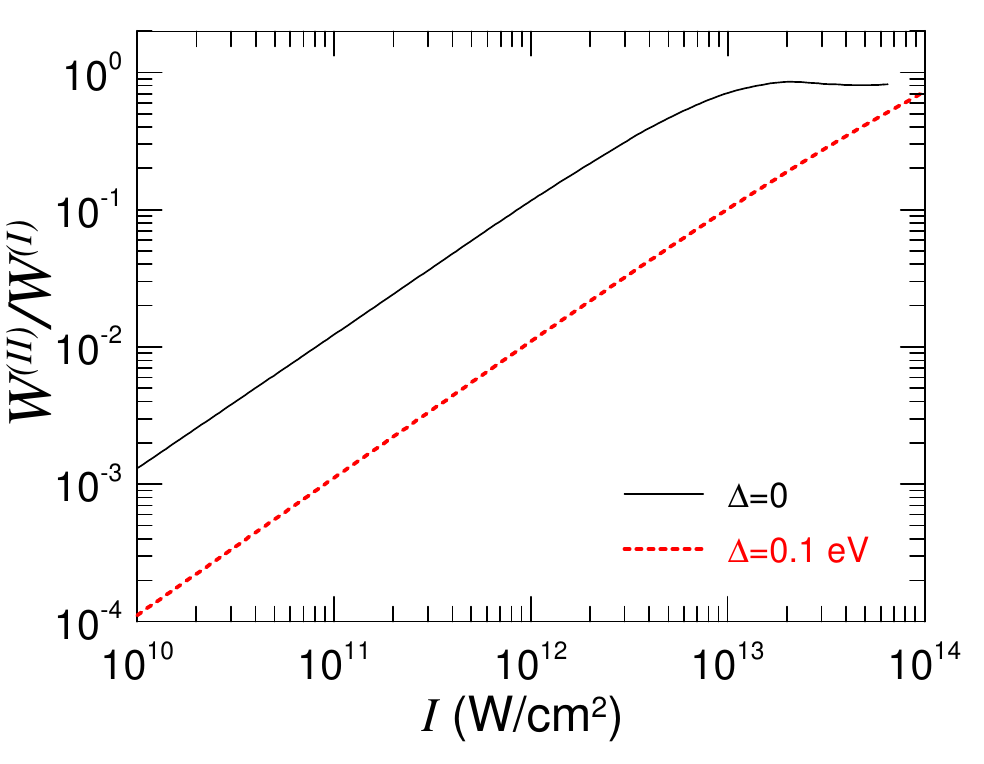}
 \caption{Ratio of signals (angle-integrated) produced by the fundamental ($W^{II}$) and the
 second harmonic ($W^{II}$) as function of
 the peak intensity. The fundamental frequency is at (detuning $\Delta = 0$) or close to ($\Delta = 0.1$~eV)
 resonant excitation of the $4s'$ state.  The other pulse parameters are
 $N=500$ and $\eta=0.1$.}
  \label{fig:fig6}
\end{figure}

Furthermore, ionization of the system becomes saturated at high intensities~\cite{Nikolopoulos2015}.
As seen in Fig.~\ref{fig:fig6}, until a peak intensity of about $10^{13}$~W/cm$^2$, our TDSE calculations predict the
angle-integrated ratio $W^{(II)}/W^{(I)}$ (fundamental over second harmonic) to
increase in a similar way with increasing peak intensity without ($\Delta = 0$) and with
small ($\Delta = 0.1$~eV) detuning, with the latter being about an order of magnitude smaller
than for the resonant case. Above $I=10^{13}$~W/cm$^2$, however,
the ratio without detuning exhibits a kink, and by $10^{14}$~W/cm$^2$, the detuning no longer
affects the ratio.

The above saturation effect, however, is not reached until a much higher peak intensity than
predicted in a previous paper (cf.\ Fig.~2 of~\cite{Nikolopoulos2015}),
even though the potential used in our TDSE calculation gives bound-bound matrix elements that are about a factor
of two larger than those of~\cite{Nikolopoulos2015}. We believe that the early saturation is the
result of the few-level model used in~\cite{Nikolopoulos2015}, where direct two-photon transitions
(primarily into the \hbox{$f$-wave}) were omitted. As mentioned above, and confirmed by our TDSE calculations,
these transitions are as important as the resonance transitions already for a detuning as small as $0.1$~eV.
Consequently, for the peak intensities around $10^{12}$~W/cm$^2$ considered in the present paper,
the saturation effect is not expected to play a significant role that would invalidate the principal
conclusions.

\section{Summary and Conclusions}\label{sec:ack}
We have investigated quantum coherent control of the photoelectron angular distribution in bichromatic
ionization of neon. Three theoretical approaches were applied
to analyze the contributions from different channels and to disentangle the potential and resonant parts
of the transition amplitudes.  To the extent that one could expect in light of the slightly
different atomic structure models and the different pulse lengths, good agreement between the non-relativistic
TDSE and PT$_\infty$ models was obtained.  This comparison provided confidence when applying the semi-relativistic
PT model, which could account for both a finite pulse length and relativistic effects in the target
description.  
In the future, we plan to extend the single-active-electron TDSE model to account for many-electron
effects by further developing the method described by Guan {\it et al}.~\cite{Guan2007}.

The principal results of our study are the estimates for the optimal pulse parameters needed to
effectively observe coherent control in realistic current setups such as that described in~\cite{Prince2016}.
Our calculations suggest that a minimal pulse duration (temporal coherency) of about 100 optical cycles
in the short-wave\-length regime is required to form a measurable asymmetry.  We also predicted
the optimal strength ratio between the fundamental and the second harmonic.

Finally, we investigated the possible effect of reaching the saturation regime, where the
yields of one-photon and two-photon processes become comparable, independent of the detuning between the
fundamental frequency and the intermediate resonance state. In contrast to previous
predictions~\cite{Nikolopoulos2015}, we find that the effect does not set in until peak intensities larger than
about $10^{13}$~W/cm$^2$ for the Ne$(4s')$ state under consideration, and hence it is not
expected to significantly alter our findings.

\section*{Acknowledgments}
The authors benefitted greatly from stimulating discussions
with Giuseppe Sansone, Kevin Prince, and Kiyoshi Ueda.
The work of N.D.\ and K.B.~was supported by the United States National
Science Foundation under grant \hbox{No.~PHY-1430245} and the XSEDE allocation \hbox{PHY-090031}.
The TDSE calculations were performed on Stampede at the Texas Advanced Computing Center and
SuperMIC at the Center for Computation \& Technology at Louisiana State University.

\end{document}